

\hsize=6.0truein
\vsize=8.5truein
\voffset=0.25truein
\hoffset=0.1875truein
\tolerance=1000
\hyphenpenalty=500
\def\monthintext{\ifcase\month\or January\or February\or
   March\or April\or May\or June\or July\or August\or
   September\or October\or November\or December\fi}


\font\tenrm=cmr10 scaled \magstep1   \font\tenbf=cmbx10 scaled \magstep1
\font\sevenrm=cmr7 scaled \magstep1  
\font\fiverm=cmr5 scaled \magstep1   

\font\teni=cmmi10 scaled \magstep1   \font\tensy=cmsy10 scaled \magstep1
\font\seveni=cmmi7 scaled \magstep1  \font\sevensy=cmsy7 scaled \magstep1
\font\fivei=cmmi5 scaled \magstep1   \font\fivesy=cmsy5 scaled \magstep1

\font\tentt=cmtt10 scaled \magstep1
\font\tenit=cmti10 scaled \magstep1
\font\tensl=cmsl10 scaled \magstep1

\def\twelvepoint{\def\rm{\fam0\tenrm}
   \textfont0=\tenrm \scriptfont0=\sevenrm \scriptscriptfont0=\fiverm
   \textfont1=\teni  \scriptfont1=\seveni  \scriptscriptfont1=\fivei
   \textfont2=\tensy \scriptfont2=\sevensy \scriptscriptfont2=\fivesy
   \textfont\itfam=\tenit \def\it{\fam\itfam\tenit}
   \textfont\ttfam=\tentt \def\tt{\fam\ttfam\tentt}
   \textfont\bffam=\tenbf \def\bf{\fam\bffam\tenbf}
   \textfont\slfam=\tensl \def\sl{\fam\slfam\tensl} \rm
   \hfuzz=1pt\vfuzz=1pt
   \setbox\strutbox=\hbox{\vrule height 10.2pt depth 4.2pt width 0pt}
   \parindent=24pt\parskip=1.2pt plus 1.2pt
   \topskip=12pt\maxdepth=4.8pt\jot=3.6pt
   \normalbaselineskip=14.4pt\normallineskip=1.2pt
   \normallineskiplimit=0pt\normalbaselines
   \abovedisplayskip=13pt plus 3.6pt minus 5.8pt
   \belowdisplayskip=13pt plus 3.6pt minus 5.8pt
   \abovedisplayshortskip=-1.4pt plus 3.6pt
   \belowdisplayshortskip=13pt plus 3.6pt minus 3.6pt
   \topskip=12pt \splittopskip=12pt
   \scriptspace=0.6pt\nulldelimiterspace=1.44pt\delimitershortfall=6pt
   \thinmuskip=3.6mu\medmuskip=3.6mu plus 1.2mu minus 1.2mu
   \thickmuskip=4mu plus 2mu minus 1mu
   \smallskipamount=3.6pt plus 1.2pt minus 1.2pt
   \medskipamount=7.2pt plus 2.4pt minus 2.4pt
   \bigskipamount=14.4pt plus 4.8pt minus 4.8pt}

\twelvepoint



\font\titlerm=cmr10 scaled \magstep3
\font\titlerms=cmr10 scaled \magstep1 
\font\titlei=cmmi10 scaled \magstep3  
\font\titleis=cmmi10 scaled \magstep1 
\font\titlesy=cmsy10 scaled \magstep3      
\font\titlesys=cmsy10 scaled \magstep1  
\font\titleit=cmti10 scaled \magstep3 
\skewchar\titlei='177 \skewchar\titleis='177 
\skewchar\titlesy='60 \skewchar\titlesys='60 

\def\titlefont{\def\rm{\fam0\titlerm}
   \textfont0=\titlerm \scriptfont0=\titlerms 
   \textfont1=\titlei  \scriptfont1=\titleis  
   \textfont2=\titlesy \scriptfont2=\titlesys 
   \textfont\itfam=\titleit \def\it{\fam\itfam\titleit} \rm}


\def\preprint#1{\baselineskip=19pt plus 0.2pt minus 0.2pt \pageno=0
   \begingroup
   \nopagenumbers\parindent=0pt\baselineskip=14.4pt\rightline{#1}}
\def\title#1{
   \vskip 0.9in plus 0.45in
   \centerline{\titlefont #1}}
\def\secondtitle#1{}
\def\author#1#2#3{\vskip 0.9in plus 0.45in
   \centerline{{\bf #1}\myfoot{#2}{#3}}\vskip 0.12in plus 0.02in}
\def\secondauthor#1#2#3{}
\def\addressline#1{\centerline{#1}}
\def\abstract{\vskip 0.7in plus 0.35in
   \centerline{\bf Abstract}
   \smallskip}
\def\finishtitlepage#1{\vskip 0.8in plus 0.4in
   \leftline{#1}\supereject\endgroup}

\def\date#1{\finishtitlepage{#1}}

\def\nolabels{\def\eqnlabel##1{}\def\eqlabel##1{}\def\figlabel##1{}%
   \def\reflabel##1{}}
\def\writelabels{\def\eqnlabel##1{%
   {\escapechar=` \hfill\rlap{\hskip.11in\string##1}}}%
   \def\eqlabel##1{{\escapechar=` \rlap{\hskip.11in\string##1}}}%
   \def\figlabel##1{\noexpand\llap{\string\string\string##1\hskip.66in}}%
   \def\reflabel##1{\noexpand\llap{\string\string\string##1\hskip.37in}}}
\nolabels


\global\newcount\secno \global\secno=0
\global\newcount\meqno \global\meqno=1

\def\newsec#1{\global\advance\secno by1
   \xdef\secsym{\the\secno.}
   \global\meqno=1\bigbreak\medskip
   \noindent{\bf\the\secno. #1}\par\nobreak\smallskip\nobreak\noindent}
\xdef\secsym{}

\def\appendix#1#2{\global\meqno=1\xdef\secsym{\hbox{#1.}}\bigbreak\medskip
\noindent{\bf Appendix #1. #2}\par\nobreak\smallskip\nobreak\noindent}


\def\eqnn#1{\xdef #1{(\secsym\the\meqno)}%
   \global\advance\meqno by1\eqnlabel#1}
\def\eqna#1{\xdef #1##1{\hbox{$(\secsym\the\meqno##1)$}}%
   \global\advance\meqno by1\eqnlabel{#1$\{\}$}}
\def\eqn#1#2{\xdef #1{(\secsym\the\meqno)}\global\advance\meqno by1%
   $$#2\eqno#1\eqlabel#1$$}


\def\myfoot#1#2{{\baselineskip=14.4pt plus 0.3pt\footnote{#1}{#2}}}
\global\newcount\ftno \global\ftno=1
\def\foot#1{{\baselineskip=14.4pt plus 0.3pt\footnote{${\the\ftno}$}{#1}}%
   \global\advance\ftno by1}


\global\newcount\refno \global\refno=1
\newwrite\rfile

\def\ref{[\the\refno]\nref}
\def\nref#1{\xdef#1{[\the\refno]}\ifnum\refno=1\immediate
   \openout\rfile=refs.tmp\fi\global\advance\refno by1\chardef\wfile=\rfile
   \immediate\write\rfile{\noexpand\item{#1\ }\reflabel{#1}\pctsign}\findarg}
\def\findarg#1#{\begingroup\obeylines\newlinechar=`\M\passarg}
   {\obeylines\gdef\passarg#1{\writeline\relax #1M\hbox{}M}%
   \gdef\writeline#1M{\expandafter\toks0\expandafter{\striprelax #1}%
   \edef\next{\the\toks0}\ifx\next\null\let\next=\endgroup\else\ifx\next\empty%

\else\immediate\write\wfile{\the\toks0}\fi\let\next=\writeline\fi\next\relax}}
   {\catcode`\%=12\xdef\pctsign{
\def\striprelax#1{}

\def\semi{;\hfil\break}
\def\addref#1{\immediate\write\rfile{\noexpand\item{}#1}} 

\def\listrefs{\vfill\eject\immediate\closeout\rfile
   \centerline{{\bf References}}\bigskip{\frenchspacing%
   \catcode`\@=11\escapechar=` %
   \input refs.tmp\vfill\eject}\nonfrenchspacing}

\def\startrefs#1{\immediate\openout\rfile=refs.tmp\refno=#1}


\global\newcount\figno \global\figno=1
\newwrite\ffile
\def\fig{\the\figno\nfig}
\def\nfig#1{\xdef#1{\the\figno}\ifnum\figno=1\immediate
   \openout\ffile=figs.tmp\fi\global\advance\figno by1\chardef\wfile=\ffile
   \immediate\write\ffile{\medskip\noexpand\item{Fig.\ #1:\ }%
   \figlabel{#1}\pctsign}\findarg}

\def\listfigs{\vfill\eject\immediate\closeout\ffile{\parindent48pt
   \baselineskip16.8pt\centerline{{\bf Figure Captions}}\medskip
   \escapechar=` \input figs.tmp\vfill\eject}}


\def\letter{\raggedright\parindent=0pt}
\def\endmode{}
\def\longindent{\parindent=3.25truein\obeylines\parskip=0pt}
\def\letterhead{\null\vfil\begingroup
   \parindent=3.25truein\obeylines
   \def\endmode{\medskip\endgroup}}

\def\sendingaddress{\endmode\begingroup
   \parindent=0pt\obeylines\def\endmode{\medskip\endgroup}}

\def\salutation{\endmode\begingroup
   \parindent=0pt\obeylines\def\endmode{\medskip\endgroup}}

\def\body{\endmode\begingroup\parskip=\smallskipamount
   \def\endmode{\medskip\endgroup}}

\def\closing{\endmode\begingroup\longindent
   \def\endmode{\endgroup}}

\def\signed{\endmode\begingroup\longindent\vskip0.8truein
   \def\endmode{\endgroup}}

\def\endofletter{\endmode \ifnum\pageno=1 \nopagenumbers\fi
   \vfil\vfil\eject\end}


\def\noblackbox{\overfullrule=0pt}
\def\inv{{\raise.18ex\hbox{${\scriptscriptstyle -}$}\kern-.06em 1}}
\def\dup{{\vphantom{1}}}
\def\Dsl{\,\raise.18ex\hbox{/}\mkern-16.2mu D} 
\def\dsl{\raise.18ex\hbox{/}\kern-.68em\partial}
\def\slash#1{\raise.18ex\hbox{/}\kern-.68em #1}
\def\lspace{}
\def\lbspace{}
\def\boxeqn#1{\vcenter{\vbox{\hrule\hbox{\vrule\kern3.6pt\vbox{\kern3.6pt
   \hbox{${\displaystyle #1}$}\kern3.6pt}\kern3.6pt\vrule}\hrule}}}
\def\mbox#1#2{\vcenter{\hrule \hbox{\vrule height#2.4in
   \kern#1.2in \vrule} \hrule}}  
\def\bar{\overline}
\def\e#1{{\rm e}{\textstyle#1}}
\def\del{\partial}
\def\curly#1{{\hbox{{$\cal #1$}}}}
\def\curlyD{\hbox{{$\cal D$}}}
\def\curlyL{\hbox{{$\cal L$}}}
\def\vev#1{\langle #1 \rangle}
\def\psibar{\overline\psi}
\def\lform{\hbox{$\sqcup$}\llap{\hbox{$\sqcap$}}}
\def\darr#1{\raise1.8ex\hbox{$\leftrightarrow$}\mkern-19.8mu #1}
\def\half{{\textstyle{1\over2}}} 
\def\roughly#1{\ \lower1.5ex\hbox{$\sim$}\mkern-22.8mu #1\,}
\def\MSbar{$\bar{{\rm MS}}$}
\hyphenation{di-men-sion di-men-sion-al di-men-sion-al-ly}
\def\l{\lambda}
\def\r{\rho}
\def\w{\omega}
\def\t{\tau}
\def\tP{\tilde P}
\def\B{{\cal B}}
\def\P{\cal P}
\def\s{\sigma}
\def\k{\kappa}
\def\a{\alpha}
\def\b{\beta}
\def\al{\alpha_{(i)}}
\def\bigsum{\sum_{l=1}{q-1}\sum_{k=1}\infty}
\def\D{{\cal D}}
\def\O{\cal O}
\def\Donetwo{\D{ij}_{1,2}}
\def\TP{\tilde P}
\def\Done{\D{ij}_1}
\def\Dtwo{\D{ij}_2}
\def\R{{\cal R}}
\def\Rlk{{\cal R}_{l,k}}
\def\Rjlk{{\cal R}j_{l,k}}
\def\Rilk{{\cal R}i_{l,k}}
\def\tlk{ t_{l,k}}
\def\rline{{{\rm I}\!{\rm R}}}
\def\pq{[P,Q]=1}
\def\pqq{[{\tilde P},Q]=Q}
\def\zplus{z\rightarrow +\infty}
\def\zminus{z\rightarrow -\infty}

\def\subsec#1{
\bigskip
\noindent
{\sl #1}
\medskip
\noindent}

\def\hf{\half}
\def\nonp{non-perturbative}
\def\hmm{hermitian matrix model}
\def\integ#1#2#3{\int_{#1}{#2}\!\! d#3\ }

\preprint{\vbox{\rightline{SHEP 90/91--30}
\vskip2pt\rightline{Imperial/TP/91-92/01}}}
\vskip -2cm
\title{\vbox{\centerline{Stable Non--Perturbative Minimal Models}
\vskip2pt\centerline{Coupled to 2D Quantum Gravity}}}
\vskip -2cm
\author{\bf Clifford Johnson$1$, Tim Morris$1$ and Bill Spence$2$}{}{}
\addressline{\it $1$Physics Department}
\addressline{\it Southampton University}
\addressline{\it Southampton, SO9 5NH}
\addressline{\it U.K.}
\addressline{}
\addressline{\it $2$Blackett Laboratory}
\addressline{\it Imperial College}
\addressline{\it London, SW7 2BZ}
\addressline{\it U.K.}

\vskip -1cm


\abstract
A generalisation of the non--perturbatively stable solutions of
string equations which respect the KdV flows, obtained recently
for the $(2m-1,2)$ conformal minimal models coupled to
two--dimensional quantum gravity, is presented for the $(p,q)$ models.
These string equations are the most general string equations
compatible with the $q$--th generalised KdV flows.
They exhibit a close relationship with the bi-hamiltonian
structure in these hierarchies.
The Ising model is studied as a particular example, for which a real
non-singular numerical
solution to the string susceptibility is presented.
\date{February 1992}

\def\nuke{Nucl.Phys.}
\def\pl{Phys.Lett.}
\nref\cmm{S.Dalley, C.V.Johnson and T.R.Morris, \nuke\ {\bf B368} (1992) 625.}
\nref\npqg{S.Dalley, C.V.Johnson and T.R.Morris, \nuke\ {\bf B368} (1992) 655.}
\nref\npqga{S.Dalley, C.V.Johnson and T.R.Morris, \nuke\ {\bf B} (Proc. Suppl.)
{\bf 25A} (1992) 87, Proceedings of the workshop on {\it Random
Surfaces and 2D Quantum Gravity,} Barcelona 10--14 June 1991.}

\nref\simon{S.Dalley, Princeton University Preprint PUPT--1290, to appear in
Mod. Phys. Lett. {\bf A}}

\nref\moo{G.Moore, M.Plesser and S.Ramgoolam, Yale preprint YCTP--P35--91.}
\nref\three{
E.Br\'{e}zin and V.Kazakov, Phys.Lett. {\bf B236} (1990) 144\semi
M.Douglas and S.H.Shenker, Nucl.Phys. {\bf B335} (1990) 635\semi
D.J.Gross and A.A.Migdal, Phys.Rev.Lett. {\bf 64} (1990) 127\semi
D.J.Gross and A.A.Migdal, Nucl. Phys. {\bf B}340 (1990) 333.}
\nref\douglas{M.Douglas, Phys.Lett. {\bf B238} (1990) 176.}
\nref\boutroux{P.Boutroux, Ann. \'Ecole Norm. Sup\'er. {\bf 30} (1913) 255\semi
F.David, Nucl.Phys. {\bf B348} (1991) 507.}
\nref\badloop{F.David, Mod. Phys. Lett. {\bf A5} (1990) 1019.}
\nref\space{
S.Dalley, Phys.Lett. {\bf B}253 (1991) 292\semi
S.Dalley, C.V.Johnson and T.R.Morris, Mod.Phys.Lett. {\bf A6} (1991) 439.}
\nref\wallone{J.Ambj\o rn, C.V.Johnson and T.R.Morris, to appear in \nuke\ {\bf
B}.}
\nref\instantons{P.Ginsparg and J.Zinn--Justin, Phys. Lett. {\bf B255} (1991)
189.}
\nref\zinn{J.Zinn--Justin, {\it Quantum Field Theory and Critical Phenomena},
Oxford University Press, 1991.}
\nref\shyamjoe{S.Chaudhuri and J.Lykken, \nuke\ {\bf B367} (1992) 614.}
\nref\mehta{M.Mehta, Comm. Math. Phys. {\bf 79} (1981) 327.}
\nref\Itzykson{C.Itzykson and J.--B. Zuber, J. Math. Phys. {\bf 21} (1980)
411.}\

\nref\tim{T.R.Morris, \nuke\ {\bf B356} (1991) 703.}
\nref\flows{C.V.Johnson and A.W\"atterstam, in preparation.}
\nref\bound{E.Martinec, G.Moore and N.Seiberg, \pl\ {\bf B263} (1991) 190.}
\nref\Ising{
V.A.Kazakov, \pl\ {\bf A119} (1986) 140\semi
D.V.Boulatov and V.A.Kazakov, \pl\ {\bf B186} (1987) 379\semi
E. Bre\'zin, M. Douglas, V. Kazakov and S. H. Shenker, Phys. Lett. {\bf B237}
(1990) 43\semi
D.J.Gross and A.A.Migdal, Phys.Rev.Lett {\bf 64} (1990) 717\semi
C.Crnkovi\'{c}, P.Ginsparg and G.Moore, Phys.Lett. {\bf B237} (1990) 196.}
\nref\Gelfand{I.M.Gel'fand and L.A.Dikii, Russian Math. Surveys {\bf 30:5}
(1975) 77.}
\nref\tada{T.Tada and M.Yamaguchi, \pl\ {\bf B250} (1991) 38\semi
T.Tada, \pl\ {B259} (1991) 442.}
\nref\segal{G.Segal and G.Wilson, Pub.Math. I.H.E.S. {\bf 61} (1985) 5.}
\nref\drinfeld{V.G.Drinfeld and V.V.Sokolov, J.Sov.Math. {\bf 30} (1985) 1975.}
\nref\lee{M.Staudacher, Nucl.Phys. {\bf B336} (1990) 349.}
\nref\Brezin{E.Br\'{e}zin, Marinari and G.Parisi, \pl\ {\bf B242} (1990) 35.}
\nref\olver{P.J.Olver, {\it Applications of Lie Groups to Differential
Equations}, Springer--Verlag, 1986}
\nref\goeree{J.Goeree, \nuke\ {\bf B355} (1990) 737.}
\nref\kawai{M.Fukuma, H Kawai and R.Nakayama, UT--572--Tokyo, KEK--TH--272,
November 1990.}

\newsec{Introduction and Conclusions.}

Pure two dimensional quantum gravity is known to have a KdV flow symmetry to
all orders in genus perturbation theory. In a series
 of papers\cmm--\simon\ it was established that a complete formulation of
non-perturbative pure two
 dimensional quantum gravity can be developed from the single
 principle that this symmetry is respected non-perturbatively i.e. that the KdV
flows are exact. In particular the principle leads to a unique string equation
which has as special cases the
non-perturbatively sick solutions of \hmm s, but also has a
 unique\foot{We believe. In ref.\npqg\ it was proven that there is at most
 a discrete number of such solutions with real asymptotics, and a
 numerical study uncovered only one.}\ real non-singular solution. The above
discoveries are
 briefly reviewed in section 2, together with the
 matrix model reasons for expecting these successes. The primary
reason for the present paper is to show that these successes
generalise to 2D quantum gravity coupled  to a general conformal
minimal model. For $(p,q)$ matter the single principle is that the $q$-th
generalised KdV flows (known to exist
 perturbatively\douglas )
are preserved non-perturbatively. We will demonstrate that this leads to unique
string equations for each system as
conjectured in ref.\npqg\ (and indeed for
 massive theories interpolating between all $(p,q)$ critical
points for given $q$) and display a unique\foot{Subject to similar caveats.}\
real singularity-free solution for the Ising model coupled to gravity.

The success of the present formulation demands an analysis of its
principle: non-perturbative preservation of flows. Let us first
note that if this principle is discarded then other formulations
are possible\wallone ; Clearly {\sl some} input is needed to define 2D gravity
(a.k.a. 1D string theory)
beyond the genus expansion. One might hope that unitarity ensures a unique
non-perturbative extension. For 2D string theory (the lowest dimension where
the concept of an
 $S$-matrix makes sense) this appears to be
 insufficient\moo . In this case it is a
 natural conjecture that the perturbation expansion has a
form of
KP flow symmetry and it would certainly be interesting to
trace this out and determine whether or not an exact KP flow symmetry
picks out a unique unitary non-perturbative extension.
Of course some other general constraint (causality?) conceivably
might provide the missing information but this is not really the
 point: If one succeeds in providing an `unprincipled' extension
(in the sense that the perturbation theory must be separately
 determined) then non-perturbative string theory remains
logically incomplete.

Having argued for principles, what form should they take? A
 common attempt in the past has been to try to formulate
 non-perturbative string theory from some symmetry principle
 on the world-sheet i.e. a symmetry of 2D quantum gravity
 considered on a single (possibly pinched) genus. In
view of the work of the last two years on low dimensional string
 theory, and of the simplicity in this case of the world-sheet theory when
 appropriately formulated, this now surely seems as unlikely
as expecting, say, non-perturbative QCD to arise from some
symmetry of the world-line.
Indeed in a second quantized theory one expects the symmetries
 to be best manifested on the second quantized fields.
Recalling that their expectation values -- the
background fields -- are nothing but the couplings $t_r$ in string theory,  we
expect the symmetry to be manifested as active transformations  on the $t_r$.
Thus a non-perturbative string
 theory symmetry principle should be a symmetry of ``theory space'', the
space of all world-sheet couplings, and not of a single
 world-sheet theory. This is precisely what the KP (generalized KdV) flows are.
 Thus it seems highly probable
that this symmetry principle is a hint of
 a much larger symmetry determining the non-perturbative form
of 2D quantum gravity coupled to general conformal matter.

In this paper our primary purpose, as already mentioned, is to develop the
formulation for gravity coupled to $(p,q)$ matter. The first steps towards this
were already taken in ref.\npqg\
where, motivated by a matrix model whose eigen-value space was
$\rline_+$, use was made  of $[\tP,Q]=Q$: the appropriate generalisation of
Douglas'
$[P,Q]=1$ formalism\douglas . By
using the Lax pair formalism for (generalized) KdV flows
we will see that in this case the Douglas' formalism  is nothing
but a trivial expression of scaling: i.e. it follows directly
from
the fact that KP flows have a grading. Many other intimate
connections with the reductions of the KP hierarchy are
uncovered. For example the classical $W{(q)}$-algebra of the
second hamiltonian structure is seen to play a central r\^ole.
The integrability of the hierarchy implies the existence of
a first integral of the scaling equation: our string equation.
The existence of a natural `gauge' parameter $\sigma$ -- the boundary of
$\rline_+$ and
physically a world-sheet boundary coupling -- is seen to be
a consequence of coordination of the bi-hamiltonian structure.
Indeed another reason for the present paper is to provide a
complete discussion of this parameter, partial results
having been reported earlier\npqga\wallone\simon .

In section 5 we recall\npqg\ that the $L_{-1}$ symmetry
of KP flows is not a symmetry of the vacuum,
leading to an analogy with spontaneous symmetry breaking
in which $\sigma$ is identified with the Goldstone boson.
Since, as we will show, all the $W{(k)}_n$ generators with
negative index $n$ are `spontaneously broken' there are
further generalizations of the $\sigma$ parameter for the $(p,q)$
models with $q\ge3$. For pure gravity $\sigma$ is the boundary
cosmological constant. For the Ising model it is the boundary
magnetic field. There are two other parameters for the Ising
model associated with $W_{-1}$ and $W_{-2}$. The latter
 we tentatively associate with the boundary cosmological
constant -- a parameter missing in previous
 formulations\bound .

The structure of the paper is as follows: Section 2 is a short
review of previous work followed by a review of our construction
for the $(2m-1,2)$ models. We  emphasise the Lax pair formulation and bring
together the earlier results on $\sigma$.

Section 3 deals with the Ising model coupled to 2D quantum
 gravity. In particular we review the $[P,Q]=1$ equations
and combine the reasons for expecting the solutions to have similar
non-perturbative sicknesses to the $[P,Q]=1$ solutions of pure 2D quantum
gravity. Next we derive the most general string equation
compatible with an exact Boussinesque flow symmetry (assuming
no new dimensionful parameters appear at the non-perturbative
level).
We study the solution
 for vanishing magnetic field in particular. The
cosmological constant $z\to+\infty$ limit is fixed by
the  established genus expansion. In the $z\to-\infty$
limit we assume that all singularity-free solutions
have an asymptotic expansion in which case there are
two possibilities. One of these leads to the
problematic $[P,Q]=1$ solution, while the other has $\rho\to0$ as
previously\npqg . There are at most a discrete number
of solutions with the latter asymptotic, and a numerical study reveals
only one. It is real and free of singularities. Since
our Lee-Yang (5,2) solution has the same leading asymptotics
we display it too, for comparison.

In section 4 we construct the most general string
equations compatible with the $q$th KdV flows. As mentioned above we generalise
the introduction of the $\sigma$ parameter, utilise the close relationship to
the generalized hierarchies, and in sect. 5 discuss the modifications
of the Dyson-Schwinger $W$-algebra constraints.

\newsec{Review}
In this section we review the $\pqq$ formulation of the $(2m-1,2)$ models
which was developed in \cmm\npqg\npqga, including a complete discussion
of the r\^ole played by the non--perturbative parameter $\s$, the boundary
cosmological constant. Central to the discussion is the
requirement of scaling in the models and the principle that the
KdV flows are preserved. We derive
the most general string equations compatible with these requirements, using the
Lax pair representation of the KdV flows. This representation readily makes
contact
with Douglas' differential operator formulation of one--matrix models,
and prepares the way for the generalisations presented
in later sections. These equations have been shown to have real, pole--free
solutions \cmm\npqg\flows.

\subsec{2.1 Matrix Models}
The original one--hermitian matrix models \three\ provided an exact solution
to the $(2m-1,2)$ models via the string equations for the string
susceptibility,
$\r$, together with the KdV flows. The solutions for $\r$ obtained from these
string equations, although well defined in perturbation theory, produce
problematic
non--perturbative solutions for the $m$--even models: The only relevant real
solutions to the string equations possess poles. The physical interpretation
of these poles is unclear, and their presence violates the Dyson--Schwinger
equations of the models. The relevant pole--free solutions to the $m$--even
equations are the triply truncated solution of Boutroux and its
generalisations\boutroux\shyamjoe\ which are complex and therefore physically
unacceptable.

The problems of the definition may be traced to an instability of the
one--hermitian matrix model at its $m$--even critical points, as a careful
study of the associated scaled eigenvalue problem reveals\space.
A complementary study of the asymptotic behaviour of the $m$--even string
equations
reveals the presence of real `instanton' solutions in the single--well
eigenvalue
problem. The presence of the instantons is {\sl not} by itself a signature of
instability. However, the local topology of the eigenvalue space and the form
of the
effective potential for the scaled eigenvalues demonstrates that the definition
of the
$m$--even critical points is unstable to eigenvalues tunneling
into a different configuration.

Later, by studying one--complex matrix models\tim\cmm\ an alternative exact
solution to
the $(2m-1,2)$ models was constructed. A different set of string equations
which have
the same perturbation theory as the previous definition was found. These
equations
possess real pole--free solutions however, thereby providing a more
satisfactory
non--perturbative definition. The KdV flows of the earlier definition
assume a central r\^ole in these models, since it turns out
that the equations are the most general consistent with this structure.

The stability of these models may be traced
back to the local topology of the scaled eigenvalues: The one--complex matrix
models
studied were formulated in terms of the combination $M\dagger M$ and the new
solutions had positive definite scaled
eigenvalues. The local topology of
the critical theory is thus $\rline_+$ in contrast to the $\rline$ of the
original
one--hermitian matrix models. The resulting `infinite wall'
at the boundary
has the effect of removing the eigenvalue tunneling problem.
The same effect may be obtained simply by imposing an $\rline_+$
topology on scaled eigenvalue space.

The differential of the string equations of the original one--hermitian matrix
model
definition of the $(2m-1,2)$ models may be written as the canonical commutation
relation
between the operator $Q$ representing position $\l_s$ and the operator $P$
representing momentum ${d\over d\l_s}$: $\pq$.
The operators $P$ and $Q$ are differential operators\douglas\ in the
scaled parameter $z$.\foot{For unitary models,
$z$ is the cosmological constant and hence couples to the
puncture operator $\P$.}\  In particular, $Q=d2+u_2$ where $u_2=-\r$ and
$d\equiv\partial/\partial z$.

For a one--matrix model defined on a half--line (i.e. the scaled eigenvalue
space
has topology $\rline_+$) we may write the differentiated string equations as
the canonical commutation
relation between $Q$ and the relevant conjugate momentum, $\tilde P$, which is
scale transformations ${\l_s{d\over d\l_s}}$ about the wall at $\l_s=0$:
$\pqq$. This represents a sort of gauge fixed version.
For full generality we should introduce another parameter into the theory: the
scaled position, $\s$,
of the wall. Thus the canonical momentum is $(\l_s-\s){d\over
d\l_s}\equiv{\tilde P}-\s P$.
We thus have:
\eqn\deform{[\tP-\s P,Q]=Q-\s}
which represents the differentiated string equations of the stable
definition of the $(2m-1,2)$ models.

\subsec{2.2 The $\pqq$ Definition of the $(2m-1,2)$ Models}
In Douglas' differential operator prescription for the $(2m-1,2)$ series, local
operators in the theory are constructed via fractional powers of $Q$, the
coordinate operator\foot{See section
4 for a brief review of the fractional powers of differential operators.},
giving the infinite set
${\cal O}_k\sim Q{k+\half}_+$, $k=1,2,\ldots$  and their dimensionful
couplings $t_k$.
Operator insertions are structured according to
the KdV flows:
\eqn\kdV{{\partial Q\over\partial t_k}=\k[Q_+{k+{1\over 2}},Q]}
Here $t_0$ and the cosmological constant
$z$ are seen to be related by the non--universal normalisation $\k$ by setting
$k$ to zero in the above: $\k t_0=z$.
Thus in the unitary model ($(3,2)\equiv$ pure gravity) where $z$ is the
cosmological
constant, we have $Q\half_+\sim\P$ the puncture operator.
In what follows, we shall normalise the KdV flows with $\k=-1$.

The string equations realising the commutation relations \deform\ may be
derived
using the principle that {\sl the KdV flows are preserved }even beyond
perturbation theory.

The scale transformation operator ${\tilde P}$ is simply
\eqn\ptilde{{\tilde P}=\sum_{k=0}\infty(k+{1\over 2})t_k{\partial\over\partial
t_k}}
where, setting the scaling dimension of $Q$ to 1, the dimensions of the $t_k$
may be derived from \kdV. Using \ptilde\ and \kdV, the translation operator
follows as
$$P=\sum_{k=1}\infty(k+{1\over 2})t_k{\partial\over\partial t_{k-1}}$$
Using these definitions and \deform\ the differentiated string equation
for the $(2m-1,2)$ models is:
\eqn\stringtwo{[\sum_{k=1}\infty (k+{1\over 2})t_k Q_+{k+{1\over 2}}
-\s\sum_{k=1}\infty (k+{1\over 2})t_kQ_+{k-{1\over 2}}-{z\over 2}d,Q]=Q-\s}
where $Q=d2+u_2$. Using the fact that $[Q_+{k+{1\over
2}},Q]=\R{'}[-u_2]_{k+1}$
equation \stringtwo\ is a differential equation for $u_2$:
\eqn\stringthree
{{1\over 4}{\cal R}{'''}+(u_2-\s){\cal R}{'}+{1\over 2}u{'}_2{\cal R}=0}
where
$${\cal R}=\sum_{k=1}\infty(k+{1\over 2})t_k{\cal R}_k-z$$
The ${\cal R}_k$'s are the
Gel'fand--Dikii differential polynomials\foot{In the original work
\Gelfand\ the differential polynomials $R_k[u]$ are
normalised such that $4R_k=\R_k$.}\ in $-u_2$. In the above, we have used the
recursion
relation ${\cal D}_1\R_{k+1}={\cal D}_2\R_k$ where
${\cal D}_1\equiv d$ and ${\cal D}_2\equiv{1\over 4}d3+u_2d+{1\over
2}u{'}_2$.
The requirement for them to vanish at $u_2=0$ fixes them uniquely up to
the normalisation $\R_0$, which we set to 2.
When multiplied by ${\cal R}$ \stringthree\ may be once integrated to give:
\eqn\smiley{(u_2-\s){\cal R}2+{1\over 2}{\cal R}{\cal R}{''}-{1\over 4}({\cal
R}{'})2=0}
where the matrix model tells us to fix the constant of integration to zero
by the requirement that in the $z\rightarrow +\infty$ limit we must have the
asymptotic expansion of ${\cal R}=0$ coinciding with the hermitian matrix model
perturbative physics.
It is easily verified that the string equation \smiley\  has the generalised
Galilean transformations
\eqn\galileo{\eqalign{
 u_2 &\rightarrow u_2+\epsilon \cr
\s &\rightarrow \s+\epsilon\cr
z &\rightarrow z+\epsilon{3\over 2}t_1 \cr
 t_k &\rightarrow t_k -\epsilon(k+{3\over 2})t_{k+1} \hskip 2.5cm k \ge 0
}}
as a symmetry. Using this symmetry we may perform a redefinition of the $t_k$'s
in order to set $\s$ to zero. This corresponds to putting the potential wall
in the matrix model
at the origin of eigenvalue space. The resulting string equation was discussed
in refs.\cmm\ and \npqg\ where
it was argued
that it possesses real, pole--free solutions for the $m$th model with the
asymptotics $u_2\rightarrow z{1/m}$ (0) in the $z\rightarrow+\infty$
($-\infty$) limits.
A numerical solution for pure gravity ($m=2$) is presented in \npqg.
Further analytical and numerical study has demonstrated the consistency of
KdV differential flows between all the $m$--critical models, and the $m=1$
and $m=3$ solutions were displayed \flows.
The symmetry \galileo\ may be rewritten as a ``flow'' for
$u_2$ under the parameter $\s$:
\eqn\sigflow{
{\partial u_2\over\partial\s}=1+\sum_{k=0}\infty(k+{3\over 2})
t_{k+1}{\partial u_2\over\partial t_k}=-\R{'}
}
where we have used the KdV flows for $u_2$:
$\partial_{t_k}u_2=-\R{'}_{k+1}$
in the last step.
Equation \sigflow\ is consistent with the interpretation of the differentiated
string equation \stringtwo\ or \stringthree\ as a scale--invariance equation
for
$u_2$:
\eqn\scaleinv{\sum_{k=1}\infty(k+{1\over 2})t_k{\partial u_2\over\partial t_k}
+{z\over 2}{\partial u_2\over\partial z}+\s{\partial u_2\over\partial\s}+u_2=0}
Equations \sigflow\ and \scaleinv\ may be written as constraints on the
partition function of the theory, which plays the r\^ole of a
tau function, $\t$,
of the KdV hierarchy\foot{$\t$ and $u_2$ are related  by
$u_2=2d2\ln\t$}.
They are then
seen to be the familiar $L_{-1}$ and $L_0$ Virasoro constraints, but with
modifications in the
presence of arbitrary $\s$:
\eqn\constr{\eqalign{
L_{-1}\t&\equiv\sum_{k=1}\infty(k+{1\over 2})t_k{\partial\t\over\partial
t_{k-1}}+{1\over 4}t2_0\t=
{\partial\t\over\partial\s}\cr
L_0\t&\equiv\sum_{k=0}\infty(k+{1\over 2})t_k{\partial\t\over\partial
t_k}+{1\over 16}\t=
\s{\partial\t\over\partial\s}
}}
The rest of the Virasoro constraints are similarly modified:
\eqn\vircon{L_n\t\equiv\sum_{k=0}\infty(k+{1\over
2})t_k{\partial\t\over\partial t_{k+n}}
+{1\over 4}\sum_{k=1}n{\partial2\t\over\partial t_{k-1}\partial t_{n-k}}=
\s{n+1}{\partial\t\over\partial\s}\hskip 1cm n\ge 1
}
That $\s$ produces a term of the form $\s{n+1}\partial/\partial\s$ in the
$L_n$ is consistent with the fact that the Virasoro constraints represent
diffeomorphisms in the space of eigenvalues of the one--matrix model.
The terms in $\s$ then arise naturally as boundary terms from infinitesimal
variations
of the position of the ``wall'' at $\l_s=\s$.

It was also noted in ref.\npqga\ that  $\s$
plays the role of a boundary cosmological constant in the theory.
It takes on the mantle of the combination $t_{m-1}/(m+\half)t_m$ which was
identified
as such for the $m$--critical $\pq$ theory to first order in $t_k,k<m$\bound.
Here the identification is not perturbative. It is also cleaner and more
natural since $\s$ couples directly to $L_{-1}$, which through \galileo\
is seen to lead to an $\e{-\s\ell}$ dependence for the macroscopic loop
$\w(\ell)\sim<\e{\ell(u_2+d2)}>$, identifying $\s$ as the boundary
cosmological
constant. Thus $L_{-1}$ is the conjugate i.e. boundary length operator---as
is already obvious in the one--hermitian matrix model from the corresponding
Ward identity\bound.
We are taking consistently the one--hermitian matrix model definition for
a macroscopic loop; these issues and the full loop equations are discussed in
depth
in ref.\simon. The above observations lead naturally to generalisations when
we consider the general $(p,q)$ model. These are discussed, together with a
derivation of \vircon\ and \constr\ through the Dyson--Schwinger equations, in
section 5.


\newsec{The Ising model}
In preparation for the generalisation of the $\pqq$ formulation to all of
the $(p,q)$ models, we study the Ising model (4,3). We derive the most
general string equation compatible with the Boussinesque flows,
but postpone the now subtle issue of boundary parameters until sections 4 and
5.
We begin with a brief review of the $\pq$ formulation of the Ising model\Ising.

\subsec{3.1 $\pq$ Ising Model}
Consider the following two--hermitian matrix partition function:
\eqn\twomatrix{Z(H,g,c)=\int\!\! {\cal D}M_+{\cal
D}M_-\exp\left({-{N\over\gamma}S(M_+,M_-,H,g,c)}\right)}
where
\eqn\potentl{S(M_+,M_-,H,g,c)={\rm
Tr}(M_+2+M_-2-2cM_+M_--ge{H}M_+4-ge{-H}M_-4)}
This defines the Ising model on a random surface where $M_\pm$ represent the
two
Ising spin states at the vertices of the diagrams in the usual large--$N$
expansion
of \twomatrix. $H$ is the magnetic field.

Following refs.\mehta\ and \Itzykson\ the partition
function may be expressed in terms of the norms $h_n$ of certain polynomials
${\cal P}_n\pm$ which are orthogonal
with respect to a measure weighted by the potential \potentl:
\eqn\orthog{\int\!\! d\lambda d\mu\  \e{-{N\over\gamma}S(\lambda,\mu)}{\cal
P}_n+(\lambda){\cal P}_m-(\mu)=h_n\delta_{nm}
}
The ${\cal P}_n\pm$ satisfy a recursion relation:
\eqn\recurr{
\lambda{\cal P}_n\pm(\lambda)={\cal P}_{n+1}\pm(\lambda)+R_n\pm{\cal
P}_{n-1}\pm(\lambda)
+S_n\pm{\cal P}_{n-3}\pm(\lambda)
}
The string equations arise as the double scaling limit \Ising\ of identities
derived using \recurr\ and \twomatrix.
The scaling functions $u_2$ and $u_3$  (related to the string susceptibility
and
the magnetisation), together with the  variables $\nu$, $\mu$ and $B$ (the
physical
string coupling, cosmological constant and magnetic field) arise as the scaling
parts
of the quantities in the recursion relation.
They are related to the free energy $\Gamma$ as follows:
$u_2=-{3/2}\nu2\partial2_\mu\Gamma$ and
$u_3=\nu2\partial_\mu\partial_B\Gamma$.
Henceforth we shall absorb $\nu$ into the quantities $\mu$ and $B$ defining
$z=\mu/\nu$ and $\B=B/\nu$.

Another approach to derive the string equations is to study the double scaled
limits, $P$ and $Q$,
of operators $P_{mn}$ and $Q_{mn}$ defined by:
$${Q_{mn}\equiv\int\!\! d\lambda d\mu \,\e{-{N\over\gamma}S(\lambda,\mu)}
{{\cal P}_m+(\lambda)\over \sqrt h_m}\mu{{\cal P}_n-(\mu)\over \sqrt h_n}}$$
and
$${P_{mn}\equiv\int\!\! d\lambda d\mu \,\e{-{N\over\gamma}S(\lambda,\mu)}
{{\cal P}_m+(\lambda)\over \sqrt h_m}{d\over d\mu}{{\cal P}_n-(\mu)\over
\sqrt h_n}
}$$
The string equations then
arise from the requirement that $P$ and $Q$, which are differential operators
in $z$,
satisfy $[P,Q]=1$. This approach was first proposed in \douglas\
and carried out explicitly for the two--matrix model in
\tada\ to yield the string equations for the
Ising model.\foot{The authors in ref.\tada\ also studied potentials of higher
order than quartic
to discover that the two--matrix model can be tuned to critical points {\sl
other} than
the $(*,3)$ models. This issue will not concern us at present.}

In the double scaled limit, the operator $Q$ becomes a third order differential
operator in $z$:
$$Q=d3+{3\over 4}\{u_2,d\}+u_3$$
Here, $d$ denotes $d/dz$.
The critical point for the Ising model is realised when $P$ is a fourth order
differential operator. The requirement that $P$ and $Q$ satisfy $[P,Q]=1$
fixes $P$ to be $Q_+{4/3}$ where the `+' denotes the differential operator
part
of the pseudo--differential\foot{These objects are briefly reviewed in section
4.
The reader
is referred to one of many fine works on the calculus of pseudo--differential
operators such as ref.\segal\drinfeld\
for a comprehensive treatment
of the subject.}\ operator $Q{4/3}$.

A simple calculation yields the following
string equations:
\eqn\Isingtwo{
\eqalign{
-{1\over 12}(u{''''}_2+9u_2u{''}_2+{9\over 2}(u_2{'})2+6u_23-8u_32)&=z\cr
{2\over 3}(u{''}_3+3u_2u_3)&={\cal B}
}
}
where we have integrated once with respect to $z$. One integration
constant has been absorbed into $z$, and the other is identified with the
magnetic field
${\cal B}$ \douglas.
(To match some of the conventions used in refs.\Ising\ we must redefine
$z\rightarrow z/2$, $\nu\rightarrow \nu/\sqrt{6}$ and $u_3\rightarrow-u_3$
and recall that the string susceptibility $\r$ is $-u_2$.)
We may study some of the tree level physics obtained from these equations
by taking the leading behaviour for large cosmological constant $z$.
This amounts to neglecting all derivatives:
\eqn\tree{\eqalign{-{1\over 2} u3_2+{2\over 3}u2_3&=z\cr 2u_2u_3&=\B}}
Eliminating $u_3$ from the resulting
equations, and adopting the above conventions, we have the following
equation for the string susceptibility at the sphere level:
\eqn\sphere{\rho3+{1\over 3}{{\cal B}2\over\rho2}=z}
{}From this we may derive expressions for tree level correlators of the
puncture
operator $\cal P$ and the spin operator $\cal S$.

The Ising model \twomatrix\ with $H\neq 0$ is not symmetric under the
interchange of
$M_+$ and $M_-$. Such an interchange translates into $d\rightarrow -d$ in the
differential
operator formalism\Ising\ and this is the origin of the $Z_2$--odd
transformation
properties of ${\cal B}$ and $u_3$.
 Studying the  $Z_2$ symmetric
sector by setting $\cal B$ and $u_3$ to zero, we have the string equation for
the
string susceptibility:
\eqn\Isingthree{{2\over 27}\r{''''}-{1\over 2}(\r{'})2-\r\r{''}+\r3=z}
Large $z$ expansion supplies the genus perturbation theory
which is uniquely determined from the the spherical contribution
$\r(z)=z{1/3}$.

Equation \Isingthree\ is very  similar to the string equation for the Lee--Yang
singularity
(the $(5,2)$ model \lee) where instead the coefficient of $\r{''''}$
is $1/10$. A numerical solution to that equation with the  asymptotics
$\r\rightarrow\pm z{1/3}$
for $z\rightarrow\pm\infty$ was presented in \Brezin.
The family of $(2m-1,2)$ string equations was studied in ref.\shyamjoe\
in order to construct the solutions analogous to the `triply truncated'
solutions found by Boutroux for Painlev\'{e}$^I$. In that work, the authors
demonstrated that the generalised truncated solutions were {\sl real} for the
$m$--odd models and {\sl complex} for the $m$--even models. The Lee--Yang
string
equation falls into the former category and the Painlev\'{e}$^I$ equation
into the latter.

It is a simple matter to repeat the analysis for the Ising equation
\Isingthree.
The reality of the truncated solutions is determined by the nature of the
solutions
to the associated polynomial\foot{We refer the reader to ref.\shyamjoe\ for
details.}\
${s2-As+3A=0}$,
where $1/A=2/27$ for the Ising model and $1/10$ for Lee--Yang.
If $s$ is complex then the truncated solutions will be real, and for real $s$
they are complex.
We see that $s$ is complex for $A>1/12$ and so the Lee--Yang solution is real
($s=5\pm i\sqrt{5}$) and the Ising model solution ($s=27,9/2$) is complex.

The alert reader will note that the same polynomial determines the reality of
the
`instantons' in the asymptotic expansion of \Isingthree\ when addressing
the question of Borel resummability \instantons\zinn.
These `instanton' solutions are  merely the leading exponential
corrections to the perturbation expansion  as
a representation of the full non--perturbative solution.
The real instantons and corresponding non--resummability
for the Ising model is purely a consequence of the unitarity of the theory
(whereby
all terms in the genus expansion  contribute with positive sign) and
the typical $(2n)!$ growth of the perturbative series.

Later we will find that the $[\tilde{P},Q]=Q$ formulation supplies the
same genus expansion as for \Isingthree\ and hence the same resummation
properties
for positive cosmological constant.
Nevertheless, we shall explicitly have a {\sl real} non--perturbative solution.
This situation was already
discussed for the $(2m-1,2)$ models in \npqg.


\subsec{3.2 $[\tilde{P},Q]=Q$ Ising Model.}
We must first recall the underlying structure which exists perturbatively for
the $[P,Q]=1$ definition of the $(*,3)$ models, the
Boussinesque hierarchy, which defines the flows of $Q=d3+(3/4)\{u_2,d\}+u_3$:
\eqn\qflow{{\partial Q\over\partial t_{l,k}}=\k[Q{3k+l\over 3}_+,Q]\hskip 2cm
l=1,2\,;\hskip 1cm k=0,1\ldots\infty}
The $t_{l,k}$ are an infinite set of parameters which parametrize the
hamiltonian
flows of $Q$. $\k$ is a non--universal normalisation parameter.
We may write the equation \qflow\ as a pair of equations for
$u_2$ and $u_3$:
\eqn\qflowtwo{\alpha_{(i)}
{\partial u_i\over\partial\tlk}=\k\Done\Rj_{l,k+1}\equiv\k\Dtwo\Rjlk \hskip
3cm i,j=2,3
}
The $\Rjlk$ are differential polynomials in $u_2$ and $u_3$,
and $\alpha_{(2)}=3/2,\alpha_{(3)}=1$. The subscript bracketed $(i)$ indicates
``no sum
on $i$''. The $\Rjlk$
are the generalisation of the Gel'fand--Dikii differential polynomials for the
KdV hierarchy. The equivalence of the two hamiltonian structures defined in
\qflowtwo\ implies a recurrence relation between them. Requiring them to
vanish at $u_2=u_3=0$ fixes them completely up to the normalisations
$\Ri_{l,0}$. The first few are:
$$\eqalign{
&\R2_{1,0}=3;\hskip 2cm\R3_{1,0}=0;\cr
&\R2_{2,0}=0;\hskip 2cm\R3_{2,0}=3;\cr
&\R2_{1,1}=2u_3;\hskip 1.6cm\R3_{1,1}={3\over 2}u_2;\cr
&\R2_{2,1}=-{1\over 4}(u{''}_2+3u_22);\hskip 1.6cm\R3_{2,1}=2u_3;\cr
&\R2_{1,2}=-{1\over 12}u{''''}_2-{3\over 4}u_2u{''}_2-{3\over 2}(u_2{'})2
-{1\over 2}u3_2+{4\over 3}u2_3;\hskip 0.5cm\R3_{1,2}={2\over
3}(u{''}_3+3u_3u_2);\cr
}$$
The objects $\Done$ and $\Dtwo$ define the first and second hamiltonian
structures
of the Boussinesque hierarchy. They are a shorthand notation for the
fundamental
structures defining the Poisson bracket for functionals of $u_2$ and $u_3$.
The explicit expressions for them are:
$$\eqalign{
&\D{22}_2={2\over 3}d3+{1\over 2}u_2{'}+u_2d\cr
&\D{23}_2=u_3d+{2\over 3}u_3{'}\cr
&\D{32}_2=u_3d+{1\over 3}u_3{'}\cr
&\D{33}_2=-{1\over 18}d5-{5\over 12}u_2d3-{5\over 8}u_2{'}d2
+(-{1\over 2}u_22-{3\over 8}u_2{''})d+(-{1\over 2}u_2u_2{'}-{1\over
12}u_2{'''})
\cr
}$$
and $\D{22}_1=\D{33}_1=0;\hskip 1cm\D{23}_1=\D{32}_1=d$.
{}From the equation
$$\al{\partial u_i\over\partial t_{1,0}}=\k\Done\Rj_{1,1}
\equiv\k\Dtwo\Rj_{1,0}=\al\k u{'}_i$$
we make the identification  $z=\k t_{1,0}$. In what follows we shall set
$\k=-1$.
The scaling dimension of $Q$ supplies a natural length scale in the theory.
Fixing its dimension fixes the scaling of the $t_{l,k}$. If we assign the
scaling
dimension 2 to $u_2$, we obtain $[Q]=3$, $[\tlk]=-(3k+l)$, and $[u_3]=3$.

The central assumption which leads uniquely to the string equation
is that {\sl the Boussinesque flows \qflow\ hold at the non--perturbative
level.}
We construct $\tilde P$, the generator of scale transformations in the theory,
out of
the parameters in the Ising model with magnetic field: $t_{1,2}$ (which defines
the (4,3) model), $t_{1,0}=-z$, and
$t_{2,0}\propto \B$.
$$\tilde{P}=-7t_{1,2}{\partial\over\partial t_{1,2}}-z{\partial\over\partial
z}
-2\B{\partial\over\partial\B}$$
Using $[\partial_{t_{1,2}},Q]=-[Q_+{7/3},Q]$, we have for our differentiated
string
equation:
$$\left[-7t_{1,2} Q_+{7/3}-z{\partial\over\partial z}
-2\B{\partial\over\partial\B} ,Q\right]-3Q=0$$
which is a pair of scaling equations for $u_2$ and $u_3$:
$$\al\left(-7t_{1,2}{\partial u_i\over\partial t_{1,2}}+z{\partial
u_i\over\partial z}
+2\B{\partial u_i\over\partial\B}+iu_i\right)=0$$
Identifying $\B=-2t_{2,0}$,
these equations may be succinctly written as:
\eqn\deetwo{\Dtwo\Rj=0}
where
${\R2=\R2_{1,2}-z}$ and $\R3=\R3_{1,2}-\B$
and we have set $t_{1,2}=1/7$. It should be noted here that we may write the
derivative of the $[P,Q]=1$ string
equations \Isingtwo\ as
\eqn\deeone{\Done\Rj=0}
Finally, to obtain the string equation, we  multiply equation \deetwo\ on the
left by $\Ri$ giving us a total differential and integrate once with respect
to $z$, to give:
\eqn\strng{
\eqalign{
&{1\over 2}u_2\R_22+{2\over 3}\R_2\R{''}_2-{1\over 3}(\R{'}_2)2
+u_3\R_2\R_3
-{1\over 18}\left(\R_3\R{(4)}_3-\R{'}_3\R{'''}_3-{1\over
2}(\R{''}_3)2\right) \cr
&-{5\over 12}\left(u_2\R_3\R{''}_3-{1\over 2} u_2(\R{'}_3)2+
{1\over 2} u_2{'}\R_3\R{'}_3\right)-{1\over
12}\left(3u_22+u_2{''}\right)\R2_3=0 \cr
}
}
(For convenience of notation we have exchanged the superscripts on the $\Ri$'s
for subscripts.)
We have set the  constant of integration to zero by requiring that our
perturbative
physics obtained in the $z\rightarrow+\infty$ limit is the same  as that
obtained
from the matrix model via equations \deeone. Indeed, with the constant in
place,
the tree level string equation is:
$$(-{1\over 2} u_23+{2\over 3}u2_3-z)2-u_2(2u_2u_3-\B)={\rm constant}$$
from which we may obtain the equations \tree\ by setting each bracket and the
constant to zero.
We can then obtain the same tree level physics as the $\pq$ definition. With
the constant
set to zero we must always follow this procedure at any level of perturbation
theory to match the physics of the $\pq$ equations. This is a direct
consequence of the fact
that the structure of the string equation \strng\ always admits a $\R_2=\R_3=0$
solution.

\subsec{3.3 The Solution for the String Susceptibility.}
We now study the physics of  equation \strng\ in the absence
 of the $Z_2$ breaking quantities $u_3$ and $\B$. Setting them to zero and
adopting the
 conventions of \Isingthree\ we  have the following string equation for the
 string susceptibility, $\r$:
\eqn\Isingfour{{9\over 16}\r\R2_I-{1\over 2}\R_I\R{''}_I+{1\over
4}(\R{'}_I)2=0}
where
$$\R_I\equiv
{2\over 27}\r{''''}-{1\over 2}(\r{'})2-\r\r{''}+\r3-z$$
We now have an equation for $\r$ which is structurally identical to that
obtained for
the string susceptibility of the Lee--Yang model $(5,2)$: The coefficient
of the first term in \Isingfour\ would instead be 1, and the expression for
$\R_{LY}$
would have $1/10$ as the coefficient of the first term. The two
equations are similarly analysed.
Using dimensional arguments the asymptotic expansion for  $\zplus$
may be shown to be of the form
$$\r=z{1\over 3}\sum_{i=0}\infty A_i\left({1\over z{7\over 3}}\right)i$$
where the $A_i$'s are dimensionless constants. By substitution these can be
seen
to be determined uniquely once $A_0$ is known.
Since the same is true of the $\pq$ string equation \Isingthree\ it follows
that the resulting perturbative expansion is identical in both cases once we
have set $A_0=1$.
As in ref.\npqg\ we assume that all real solutions without
an asymptotic expansion in the $\zminus$ limit have poles. Requiring an
asymptotic expansion in this limit
we find that the sphere solution is either
$\r=-|z|{1/3}$ or $\r=0$. It is clear from the above discussion that the
first choice leads to
the same perturbative physics as the $\pq$ definition and probably the
same problematic non--perturbative solution. (Certainly
there are at most a discrete number of solutions in this
case with the latter being one of them \foot{A numerical study of the string
equation, using the techniques described later, failed to find a pole--free
solution with these asymptotics.}.)
This expectation is reinforced by the matrix model understanding of the
$(2m-1,2)$
models; in particular the Lee--Yang model's
$\pqq$ definition has the spherical physics
fixed in the one--matrix model \cmm: The end--point of the
eigenvalue density of the model corresponds to the string susceptibility $\r$
in the spherical limit.
In the $\zplus$ limit the endpoint of the density
pulls away from the wall where we recover locally the $\rline$ topology and
hence
the physics is identical to that of the $\pq$ models; $\r_0=z{1/3}$. In the
$\zminus$ limit, the endpoint pushes up against the wall, and  $\r_0$
vanishes.

The appropriateness of the choice $\r=0$ in the $\zminus$ limit for the string
equation
\Isingfour\ is made manifest by a detailed analytic study of the family of
equations of this form and their solutions\flows. The KdV and Boussinesque flow
structure of the $(2m-1,2)$ and $(*,3)$ models respectively, may be shown to
preserve
the monodromy data of the linear problem associated to the string equations
if and only if the $\r_0=0$ asymptotic is preserved.
This asymptote is fixed in pure gravity\npqg. Note that $(2,3)=(3,2)$,
a fact which is trivially verified using the expression for ${\cal
R}{2}_{2,1}$.

Using the same procedure as for the $\zplus$ limit, in the $\zminus$ limit
with $\r_0=0$ we have
the asymptotic series
$$\r={1\over z2}\sum_{j=0}\infty B_j\left({1\over z7}\right)j$$ The $B_j$'s
are again fixed uniquely by the initial  $B_0=-4/9$.

We continue by studying linear perturbations $\epsilon$ about the leading
behaviour of
the string susceptibility in the large $z$ regions. In the $\zplus$ limit we
have $\r_0=z{1/3}$
and so we try to find a solution for $\epsilon(z)$ by substituting
$\r=z{1/3}+\epsilon(z)$
into equation \Isingfour.
Following the WKB prescription for large $z$ we expect the exponential
behaviour
$\epsilon\sim \e{-f(z)}$ with $|f|>>|f{'}|>>|f{''}|\ldots$ so that
$\epsilon{(n)}/\epsilon\approx(-f{'})n$.
Using this and keeping only leading order we find that
$$\R_I(\r_0+\epsilon)\approx\epsilon\left({{2\over
27}(f{'})4-(f{'})2z{1\over 3}+3z{2\over 3}}\right)$$
and $\R{(n)}_I(\r_0+\epsilon)\approx(-f{'})n\R_I(\r_0+\epsilon)$.
(Derivatives of $\r_0$
do not survive in this limit.) The string equation \Isingfour\ becomes
$$\left({{2\over 27}(f{'})4-(f{'})2z{1\over 3}+3z{2\over 3}}\right)2
\left({9\over 4}z{1\over 3}-(f{'})2\right)=0$$
{}From this we find
$\epsilon(z)=A\e{-{6/7}\a z{7/6}}$
with $\a2=9/2,9$ or $9/4$.
As three of these solutions are exponentially growing perturbations we must set
their
coefficients to zero.

In the $\zminus$ limit we study perturbations around the leading non--vanishing
behaviour for the string susceptibility, the torus term $\r_0=-4/9z2$.
This time, as $\r_0$ and it derivatives are subleading, we have
$\R_I(\r_0+\epsilon)\approx\epsilon{2/27}(f{'})4-z$.
The string equation then becomes, to leading order
${9/16}z2+{z/27}(f{'})6=0$
which gives
$(f{'})6=-{243/16}|z|$.
This yields
the following solution for the exponential corrections in the $\zminus$ limit:
$\epsilon(z)=B\e{-{6/7}\b z{7/6}}$
with $\b6=-243/16$.
Again, three of these solutions have positive real part and their coefficients
must be set
to zero to match the chosen asymptotes.
Six integration constants  have now been determined locally
in a sixth order differential equation and so we expect at most a discrete
number of solutions with the above asymptotics.

Further progress was made with
numerical techniques identical to those used in \npqg\ to find the solution
for the $\pqq$ pure gravity model ($m=2$) and in \flows\ for the $m=1$ and
$m=3$  models.
The solution of the differential equation was treated as a two--point boundary
value problem. We employed a {\bf NAG} {\sl FORTRAN} library routine (D02RAF)
to solve the problem by relaxation. The program uses Newton iteration with
deferred
correction, and allows user specification of the initial mesh and approximate
solution. The absolute error tolerance was set at $\sim 10{-5}$.

We  chose to solve the string equation differentiated once, removing a factor
of $\R_I$. This allows for a more numerically well--behaved
highest derivative,
since otherwise the expression for the highest derivative obtained from the
string
equation contains factors of $1/\R_I$. It was ensured that the correct solution
was
found by including more terms in the asymptotic series to calculate the
boundary
conditions.
The stability of the solution
was tested by performing the integration with a number of different values for
the
boundary, e.g. $z=\pm 100$ and $z=\pm 10$.

The non--perturbative solution  for the string susceptibility of the Ising
model
is displayed in figure 1(a),
where the integration was carried out on a mesh of
1600 points in the range $\pm 200$. For contrast, in figure 1(b), we display
the non--perturbative
solution for the Lee--Yang model.

\vskip 2cm

\newsec{The $(p,q)$ String Equations and the $q$th KdV Hierarchy.}
We present here the generalisation of the $\pqq$ formulation to the $(p,q)$
minimal models, deriving the unique string equation consistent with the
requirement of preservation of
the $q$th KdV flows. A parameter analogous to $\s$ is included in the
discussion, which in the Ising model is seen to be the boundary magnetic field.

\subsec{4.1 The $q$th KdV Hierarchy.}
We begin by reviewing the basic tools of the formalism, the KP hierarchy
and  its $q$--reductions, referred to as the $q$th KdV hierarchies.
The KP hierarchy  may be formulated in terms of the pseudo differential
operator
$L=d+\sum_{i=1}\infty f_i(t)d{-i}$, where $d{-1}$ is defined by
${d{-1}f=\sum_{j=0}\infty(-1)jf{(j)}d{-j-1}}$. Integer powers of $L$
generate a basis for the complete set of objects which commute with it. Taking
the differential operator part of these (denoted by a `+' subscript) generates
a set of evolution equations for $L$:
\eqn\lflow{{\partial L\over\partial t_r}=\k[L_+r,L]}
The $t_r$ parametrise the infinite set of flows thus defined. Equation \lflow\
defines the KP hierarchy.

The {\sl $q$th reduction} may be constructed in terms of the object $Q=Lq$
and then requiring that $Q_-=0$:
\eqn\qflow{{\partial Q\over\partial t_r}=\k[Q_+{r\over q},Q]}
When $r=0$ mod $q$ the flows of equation \qflow\ are trivial,
so we modify our notation explicitly to highlight the values of $r$ which are
mutually
prime with $q$:
\eqn\gkdv{{\partial Q\over\partial\tlk}=\k[Q_+{kq+l\over q},Q]
\hskip 1.5cm l=1,2,\ldots q-1;\, k=0,1,\ldots\infty}
The indices $l$ and $k$ now span the set of non--trivial flows $r=qk+l$.
Equation \gkdv\ defines the  $q$th KdV hierarchy.
The differential operator $Q$ may be written in the form:
\eqn\qform{Q=dq + \sum_{i=2}q\alpha_i\{u_i,d{q-i}\}}
and \qflow\ defines a pair of Hamiltonian equations for the $\{u_i\}$:
\eqn\ham{{\partial u_i\over\partial\tlk}=\k\{ {\cal H}_{l,k+1},u_i\}_1
\equiv\k\{ {\cal H}_{l,k},u_i\}_2}
The hamiltonians are constructed from fractional powers of $Q$ in the following
way:
$${{\cal H}_{l,k}={q\over kq+l}\int\!\!{\rm Res}Q{kq+l\over q}dz =
{q\over kq+l}{\rm Tr}Q{kq+l\over q}}$$
where the residue of a pseudo--differential operator is simply the
coefficient of the $d{-1}$ term.
We note here that the ${\cal H}_{l,k}$ are not well defined for the relevant
solutions
$u_i$, which grow as a power of $z$ as $\zplus$. Their explicit appearance here
uncovers the structure of
the $q$--KdV system. They themselves are never used with these solutions, both
here and later.

These $q$--KdV systems are `bi--hamiltonian': They possess two Poisson brackets
between functionals $W$, $V$ of the
$\{u_i\}$:
\eqn\poisson{\{W[u],V[u]\}_{1,2}=\int\!\! dx dy
{\delta W\over\delta u_i(x)}
\{u_i(x),u_j(y)\}_{1,2}
{\delta V\over\delta u_j(y)}}
The fundamental Poisson brackets $\{u_i(x),u_j(y)\}_{1,2}$
may be written \eqn\fund{\{u_i(x),u_j(y)\}_{1,2}={\cal
D}_{1,2}{ij}(x)\delta(x-y)}
The objects $\Done$ and $\Dtwo$ are a set of differential operators.
Using them we may develop \ham\ further
\eqn\recur{\eqalign{
{\partial u_i\over\partial\tlk}&=\k\{{\cal H}_{l,k+1},u_i\}_1
=\k\{{\cal H}_{l,k},u_i\}_2\cr
&=\k{\cal D}_1{ij}{\delta{\cal H}_{l,k+1}\over\delta u_j}=
\k{\cal D}_2{ij}{\delta{\cal H}_{l,k}\over\delta u_j}\cr
&\Rightarrow {\cal D}_1{ij}\Rj_{l,k+1}={\cal D}_2{ij}\Rj_{l,k}
\hskip 3cm i,j=2,3,\ldots,q\cr
}}
where the $\Rj_{l,k}$ are differential polynomials in the $\{u_i\}$. They
are the generalisation of the Gel'fand--Dikii differential polynomials
encountered in the $q=2$ case. The last line in \recur\ is a recursion relation
among them. Requiring them to vanish at $\{u_i\}=0$ fixes them uniquely, up to
the normalisations ${\cal R}i_{l,0}\propto q \delta{i-1}_l$. In what follows
we set these at ${\cal R}i_{l,0}= q \delta{i-1}_l$, and  we set the overall
normalisation $\k$ to -1.
The second Poisson bracket in \fund\ is in fact the
$W{(q)}\equiv {\cal W}{\cal A}_{q-1}$--algebra
(at a particular value of the central charge)
where the  $\{u_i\}$
are the $q-2$ currents. In
particular, $u_2$ corresponds to the energy--momentum tensor. For example
in the $q=2$ case we have:
$$\{u_2(z),u_2(y)\}_2=\left({1\over 4}d3+u_2d+
{1\over 2}u_2{'}\right)\delta(z-y)$$
which is the $W{(2)}$ or Virasoro algebra, and for the case $q=3$ we have a
$W{(3)}$--algebra:
$$\eqalign{
&\{u_2(z),u_2(y)\}_2=\left({2\over 3}d3+{1\over
2}u_2{'}+u_2d\right)\delta(z-y)\cr
&\{u_2(z),u_3(y)\}_2=\left(u_3d+{2\over 3}u_3{'}\right)\delta(z-y)\cr
&\{u_3(z),u_2(y)\}_2=\left(u_3d+{1\over 3}u_3{'}\right)\delta(z-y)\cr
&\{u_3(z),u_3(y)\}_2=\cr
&=\left(-{1\over 18}d5-{5\over 12}u_2d3-{5\over 8}u_2{'}d2
+(-{1\over 2}u_22-{3\over 8}u_2{''})d+(-{1\over 2}u_2u_2{'}-{1\over
12}u_2{'''})
\right)\delta(z-y)\cr}$$
which forms the  second hamiltonian structure for the
Boussinesque hierarchy\olver.

\subsec{4.2 The $(p,q)$ String Equations.}
As discussed before, the matrix model formalism motivates us to work with the
operator $Q$ which plays the r\^oole  of the continuum limit of a position
operator in the orthogonal polynomial basis. We construct a realisation of the
equation $[{\tilde P}-\s P,Q]=Q-\s$ in terms of the parameters $t_{l,k}$.
Heat kernels of $Q$ generate the correlators of the observables in the theory,
which are the macroscopic loops.  $Q$ therefore supplies a natural
length scale in the theory. From the definition of $Q$ in \qform\ and the
$q$--KdV flows in \qflow\ the scaling of the $\{u_i\}$ and $\{t_{l,k}\}$ may
be deduced.
If we assign the scaling dimension 2 to $u_2$ we obtain the following
dimensions:
$[u_i]=i;[\tlk]=-(qk+l);[Q]=q$.
Using these, we construct the generator of scale transformations, $\tilde P$:
$$\TP=\sum_{l=1}{q-1}\sum_{k=0}\infty (qk+l)\tlk {\partial\over\partial\tlk}
$$
and the generator of translations, $P$:
$$P=\bigsum (qk+l)\tlk {\partial\over\partial t_{l,k-1}}$$
By then adopting the principle that the $q$th KdV hierarchy holds we have
\eqn\bigflow{[\sum_{l=1}{q-1}\sum_{k=0}\infty\k(qk+l)\tlk\left({Q{k+{l\over
q}}_+-\s Q{k+{l\over q}-1}_+}\right),Q]=q(Q-\s)}
We may rewrite this as a set of scaling equations for the $\{u_i\}$:
\eqn\bigscale{\al\left({\bigsum(qk+l)\tlk
{\partial u_i\over\partial\tlk}
+\sum_{m=1}{q-1}mt_{m,0}{\partial u_i\over\partial t_{m,0}}
+\s{\partial u_i\over\partial\s}+iu_i}\right)=0}
 The $q-1$ objects
$t_{m,0}$ are proportional to the parameters coupling to the relevant operators
in the theory,
the ${\cal O}_{m,0}$: In particular we have $t_{1,0}=-z$, $2t_{2,0}=-\B$, etc
(after
setting $\k=-1$),
and so in rewriting \bigflow\ we must use the identity
$$[z{\partial\over\partial z},Q]=\sum{q}_{i=2}\al
\left\{z{\partial u_i\over\partial z}+iu_i,d{q-i}\right\}-qQ $$
and in order to interpret it as a scaling equation we have made the
following identification:
\eqn\Galileo{\al{\partial u_i\over\partial\s}=-q\Done\Rj}
The scaling equations \bigscale\ may be written succinctly as
\eqn\Figaro{(\Dtwo-\s\Done)\Rj=0} where
the objects $\Ri$ in the above equation are
$$\eqalign{\Ri&\equiv\sum_{l=1}{q-1}\sum_{k=0}\infty(k+{l\over
q})\tlk\Rilk\cr
&=\bigsum(k+{l\over q})\tlk\Rilk+(i-1)t_{i-1,0}}$$
In the above, we have used that $\R_{l,0}j=q\delta{j-1}_l$.
Equation \Figaro\ is the differentiated string equation. Its structure
is an explicit realisation of the $W{(q)}$--algebra structure inherent
in the second Hamiltonian structure of the $q$--KdV hierarchy.
In analogy with the case explicitly worked out from the matrix model, the
string
equation is obtained by multiplying on the left by $\Ri$ giving a total
derivative, and integrating
once with respect to $z$. (The integration constant is then set to zero
using perturbation theory. See section 3.3)

It should be noted here that the $\s$--deformed differentiated string equation
for the $(2m-1,2)$
models may also be written in the form of \Figaro:
$({\cal D}_2-\s{\cal D}_1)\R=0$
It is now apparent that the process of introducing
$\s$ into the formalism may be regarded as
forming a linear combination of the second
hamiltonian structure ${\cal D}_2$ and the first, ${\cal D}_1$.
This is always possible as the two structures are `coordinated', in the
sense of ref.\drinfeld.
In this picture \Figaro\
is indeed the natural generalisation of the equations first found for
$\pqq$ definition of the $(2m-1,2)$ models.

That $\Ri(\Dtwo-\s\Done)\Rj$ is always a total derivative must be
demonstrated.
The integral $\int\!\!dz\Ri\Donetwo\Rj$ is
$$\sum_{l,m=1}{q-1}\sum_{k,n=0}\infty
(k+{l\over q})(n+{m\over q})
\tlk t_{m,n}\int\!\!\Rilk\Donetwo\Rj_{m,n}dz$$
We then have using equations \poisson\ and \fund:
$$\eqalign{\int\!\!\Rilk\Donetwo\Rj_{m,n}dz
&=\int\!\!dz\,dy\Rilk(z)\Donetwo(z)\delta(z-y)\Rj_{m,n}(y)\cr
&=\int\!\!dz\,dy{\delta{\cal H}_{l,k}\over\delta u_i(z)}
\{u_i(z),u_j(y)\}_{1,2}{\delta{\cal H}_{m,n}\over\delta u_j(y)}\cr
&=\{{\cal H}_{l,k},{\cal H}_{m,n}\}=0
}$$
where the last line is simply the statement that all the Hamiltonians in
the $q$--KdV hierarchy are in involution.
However, the above equation  holds for {\sl all} functions $u_i$ satisfying the
boundary conditions
(which are those neccessary for the Hamiltonians ${\cal H}_{l,k}$ to be
well-defined). This can only be true if the integrand on the left hand side
of the equation is a total $z$--derivative. Thus we conclude
that we may always integrate \Figaro\ to obtain the string equation.

As an example, we have the string equation for the $(*,3)$ models:
\eqn\strngtwo{
\eqalign{
&{1\over 2}u_2\R_22+{2\over 3}\R_2\R{''}_2-{1\over 3}(\R{'}_2)2
+(u_3-\s)\R_2\R_3
-{1\over 18}\left(\R_3\R{(4)}_3-\R{'}_3\R{'''}_3-{1\over
2}(\R{''}_3)2\right) \cr
&-{5\over 12}\left(u_2\R_3\R{''}_3-{1\over 2} u_2(\R{'}_3)2+
{1\over 2} u_2{'}\R_3\R{'}_3\right)-{1\over
12}\left(3u_22+u_2{''}\right)\R2_3=0 \cr
}
}
where
$$\R_2\equiv\sum_{l=1,2}\sum_{k=1}\infty(k+{l\over 3})\tlk\R2_{l,k}-z$$
and
$$\R_3\equiv\sum_{l=1,2}\sum_{k=1}\infty(k+{l\over 3})\tlk\R3_{l,k}-\B$$
(For convenience of notation we have exchanged the superscripts on the $\Ri$'s
for subscripts.) From this, the Ising model is obtained by setting
$\tlk=3/7\deltal_1\deltak_2$.

\newsec{The W--algebra Constraints.}
In this section the constraints for the $\pqq$ formulation of the $(p,q)$
models are derived.
We complete the discussion of the parameter $\s$ in the $(2m-1,2)$ models
in terms of the algebra of constraints, which provides the appropriate method
for generalisation to the $(p,q)$ models. We discuss the possible significance
of the analogous $q-2$ extra parameters.

\subsec{5.1 The $(p,q)$--Model W--constraints}
In the $\pq$ formalism, the string equation leads to an infinite set of
constraints
on the partition function of the theory, which plays the r\^ole of a
$\tau$--function
of the associated $q$th reduction of the KP hierarchy. The $\pq$ string
equation
\deeone is equivalent to the  $L_{-1}$ constraint on the $\tau$--function:
\eqn\Lminus{\Done\Rj\equiv ddi\left({L_{-1}.\tau\over\tau}\right)=0}
(For what follows we shall work with {\sl all} the $t_r$'s which parameterise
all of the flows, including the trivial ones\goeree.
The $\Ri$'s from the previous section are now defined as $\sum_{r=1}\infty
{r\over q}t_r\Ri_r$,
and in equation \Lminus\ we have $di=\partial/\partial t_{i-1}$. We also have
$u_i=(i/2) ddi(\ln\tau)$.)

In ref.\goeree\ it was shown that the constraints
derived from the string equation may be written as:
\eqn\pqw{W{(k)}_n.\tau=0,\hskip 2cm n\ge -k+1,\,\, k=2,3,\cdots,q}
where the $W{(k)}_n$ are the $n$th Fourier modes of the $W{(q)}$--algebra
generator with spin `$k$'.
For example $W{(2)}$ is the stress tensor and its Fourier modes (usually
denoted
$L_n$) satisfy the Virasoro algebra. The constraints $L_n.\tau = 0$  $n\ge -1$
then
form a consistent set in the sense that no further constraints upon $\tau$ are
generated using the commutation relations of the modes.

In this section we will show that equations \deetwo, obtained from the $\pqq$
string equation by differentiation,
imply the following
constraints:
\eqn\pqqw{W{(k)}_n.\tau=0,\hskip 2cm n\ge 0,\,\, k=2,3,\cdots,q}
which also form a consistent set.
These constraints follow from the  differentiated string equation, which is
equivalent to the $L_{0}$ constraint:
$$\Dtwo\Rj\equiv ddi\left({L_{0}.\tau\over\tau}\right)=0$$
where
$$L_0=\sum_{i=1}\infty it_i{\partial\over\partial t_i}+{\rm const.}$$
Using the techniques and notation\foot{
Beware of the interchange of the names of the operators $Q$ and $L$ in
ref.\goeree}\ of ref.\goeree, the constraint
$L_0.\tau=0$ may be written:
\eqn\resone{res_\l\left(\l\partial_\mu2X(t,\l,\mu)|_{\mu=\l}\right).\tau=0}
{}From equation (2.8) of ref.\goeree,
\eqn\restwo{\eqalign{
res_\l(
\l\omega{*}(t{'},\l)&\partial_\l\omega(t,\l))
\cr
&=res_\nu
\left(
{X(t,\nu)res_\l\left(\l\partial_\mu2X(t,\l,\mu)|_{\mu=\l}\right)\tau(t)
\over\tau(t)}
{X*(t{'},\nu)\tau(t{'})\over\tau(t{'})}
\right)
}}
and thus equation \resone\ implies that the left hand side of \restwo\
vanishes.
However, following ref.\goeree\ one has
$$res_\l\left(\l\omega{*}(t{'},\l)\partial_\l\omega(t,\l)\right)=
\left(MQ{1/q}(t)\right)_-\delta(x-x{'})$$
We have shown that the $\pqq$ string equation implies the condition
\eqn\cond{\left({MQ{1/q}}\right)_-=0.}
Taking powers of $T\equiv MQ{1/q}$, it is then straightforward to show that
equation
\cond\ implies the further conditions
$$\left({MnQ{n/q}}\right)_-=0$$
for $n=1,2,\ldots,q-1$. Following the discussion of ref.\goeree, these lead to
the
constraints \pqqw.

The above equations correspond to the case
where the scaled eigenvalue space is taken to be $\rline_+$ with
 the boundary, or `wall',  being at $\lambda_s=0$. In fact there is no reason
for this restriction on the coordinates and in
 general one may take the boundary position to be $\lambda_s=\sigma$.  The
effect of this on the string equation
 and the Virasoro constraints of the one-matrix model has already been
described in refs.\npqga\wallone\simon.
approach most suited to
 There it is shown that the eigenvalue boundary plays the r\^ole of the
boundary
 cosmological constant on the worldsheet. In sections 4.2 and 3.2, by
identifying the eigenvalue space with the position operator $Q$
 in Douglas' $PQ$ formalism, we derived the effect of $\sigma$ on the string
equation in the general $(p,q)$ case, and on the Ising model
in particular. In the latter case the $\sigma$ parameter will be seen, by
precisely parallel arguments to
 refs.\bound\simon , to be the ($Z_2$ odd) boundary magnetic field.
 This method however takes into account only one of the flavours of eigenvalue
in a multi-matrix model\foot{Strictly
 speaking it is a linear combination\bound .}, each of which may have their own
boundary.

Rather than working directly in the continuum limit the clearest
 picture might be expected to emerge from working explicitly with
 a $q-1$-matrix model. (For the one-matrix model this was done
 in ref.\simon\ where
  it was used primarily to derive the effect on  macroscopic loops.)
Integrating over the angular modes leaves an integral over $q-1$ coupled
eigenvalues, and the obvious generalisation
 is to introduce $q-1$ boundary parameters giving the position of the $q-1$
`walls' in scaled eigenvalue space.
eigenvalues restricted to being greater than their
In the Ising model, a critical point in the two-matrix model,
 linear combinations of the two flavours of loop give the $Z_2$ even boundary
length and $Z_2$
 odd boundary magnetization\bound . Given the results of the one-matrix model,
it is then natural to conjecture that here the
 wall parameters will in a similar way provide the conjugate parameters: the
boundary magnetic field and the boundary
 cosmological constant. Note that the latter is not apparent in the $\pq$ KP
description\bound . It is therefore an important
 question to determine whether such a parameter exists in our formulation. We
will below identify a $Z_2$ even parameter,
 conjugate to a redundant $Z_2$ even operator (namely $W_{-2}$) which we
therefore suggest is (perhaps non-linearly) related to
 the boundary cosmological constant. Curiously, we will also uncover a further
redundant $Z_2$ odd parameter (conjugate to
 $W_{-1}$) for which we have as yet no physical interpretation.

The reason we cannot be more definite in our identifications is because we were
unable to carry through a direct analysis
of the two-matrix model, due to
certain technical difficulties in the orthogonal polynomial approach (outlined
below).
Other technical difficulties nullify
the standard steepest descents approach as an alternative method for any
multi-matrix model.
\def\P{{\cal P}}

Let us now turn briefly to the two-matrix model inserting eigenvalue boundaries
at $\sigma_1=\sigma_2=0$. The orthogonal
 polynomials are now normalised as
$$\integ{0}\infty{\lambda d\mu}\
 \e{-{N\over\gamma}S(\lambda,\mu)}\ \P_n+(\lambda)\P-_m(\mu)
=h_n\delta_{nm}$$
The problem with the method arises when we consider the generalisation of
eqn.\recurr . The reason the recursion relation
 \recurr\ finishes at $\P_{n-3}$ is because the potential $S$ is of order 4,
as can be seen by noting that
 $\lambda\P+_n(\lambda)\P-_{n-k}(\mu)/h_{n-k}$ vanishes for any $k>3$ on
integration with the measure implied in eqn.\orthog .
 The proof follows by converting $\lambda$ into
${\gamma\over2Nc}{\partial\over\partial\mu}
 \exp(-{N\over\gamma}S)+  corrections$, integrating by parts and then using
\orthog\ together with the definition
 $\P-_m(\mu)=\mum+\ $ $lower$ $powers$.
Unfortunately in our case we pick up a boundary term
$${\gamma\over2Nc}\P-_{n-k}(0)\integ0\infty{\lambda}
\e{-{N\over\gamma}S(\lambda,0)}\ \P_n+(\lambda)$$
for any odd $k>0$ which is therefore equal to the coefficient $\P+_{n-k}$ in
an infinite order\foot{Recall that we are
 interested in $n\approx N\to\infty$ for the continuum limit.}\  generalisation
of \recurr . The coefficients (in particular the
 $h_n$) can now be determined, in principle, from an {\sl infinite} set of
simultaneous recurrence relations generalising the usual
 construction. We conclude that the orthogonal polynomial approach is at best
inappropriate  for analysing the continuum
 limit.
Of course these arguments are unaffected by choosing general positions
$\sigma_1,\sigma_2$ for the walls or using a non-even
 potential.

We now turn to an approach based on the continuum Dyson-Schwinger equations.
We first review the results of the one-matrix model
 using this approach and then generalise to general $(p,q)$.
 Our starting point in the one-matrix model is the
 algebra of constraints\npqg:
\eqn\cold{L_n\tau=0\quad\quad n\ge 0}
This differs from the \hmm\ in that the $L_{-1}$ constraint is missing. The KdV
flows on the other hand, upon which our
 formulation is based, are invariant under transformations generated by the
full set $L_n : n\ge-1$. Thus we have a
 situation reminiscent  of spontaneous symmetry breaking: the `dynamics' i.e.
the KdV flows are invariant under the full group
 (generated by the $L_n : n\ge-1$) whereas the `vacuum' $\tau$ is invariant
only under the little group
 generated by $L_n : n\ge0$. The `Goldstone boson' is $\sigma$ which in \cold\
has been gauge fixed to zero. Indeed
 there are now an infinite number of vacua $\tau(\sigma)$ connected by the
broken generator:
\eqn\taup{\tau(\sigma)=\e{\sigma L_{-1}}\ \ \tau}
where we identify $\tau(0)$ with the $\tau$ function in \cold . The constraints
on $\tau(\sigma)$ are an inner automorphism of those of
 \cold\ plus a constraint arising from $\sigma$ independence of $\tau$:
\eqnn\cinn
$$\eqalign{L'_n\tau(\sigma) &=0\quad\quad\quad\quad n\ge 0\cr
           {\partial}'_\sigma  \tau(\sigma) &=0\cr
{\rm where}\hskip 2cm {L}'_n &= \e{\sigma L_{-1}}\ L_n\  \e{-\sigma
L_{-1}}\hskip 2cm\cr
{\rm and}\hskip 2cm {\partial}'_\sigma &=
\e{\sigma L_{-1}}\  {\partial\over\partial\sigma}\
 \e{-\sigma L_{-1}}\ \ .\hskip 2cm\cr}
\eqno\cinn$$
Taking linear combinations of these constraints: 
\eqnn\linn
$$\eqalign{L\sigma_{-1} &= -{\partial}'_\sigma\cr
           L\sigma_0    &= {L}'_0 + \sigma L\sigma_{-1}\cr
           L\sigma_1    &= {L}'_1 +2\sigma {L}'_0
 +\sigma2 L\sigma_{-1}\cr
\vdots\hskip 0.2cm   &= \hskip 1.6cm\vdots}\eqno\linn$$
we get $L\sigma_n =L_n -\sigma{n+1}{\partial\over\partial\sigma}$ for
$n\ge-1$.
  These corrections can be computed most simply as follows:
\eqnn\lmone
\eqnn\direct
\def\ad{{\rm ad}}
$$\eqalignno{{\partial\over\partial\sigma}\tau(\sigma) &= L_{-1}\tau(\sigma)
&\lmone\cr
  L_n \tau(\sigma) &= \e{\sigma L_{-1}}\ \ (-\sigma \ad L_{-1}){n+1} L_n\ \
\tau(0) \quad\quad\quad\quad n\ge0\cr
&=\sigma{n+1} L_{-1} \e{\sigma L_{-1}}\ \tau(0) \cr
&= \sigma{n+1}{\partial\over\partial\sigma} \tau(\sigma) &\direct\cr}$$
where use is made of \taup , the notation $(\ad X) Y= [X,Y]$ (for two
operators), and the identity $\e{-X}\ Y\e{X} =\e{-\ad X}\ Y$.

Our string equation \smiley\ follows from integrating  the modified $L_0$
constraint. It is clear that the system is now invariant under translations
generated by $L_{-1}$. Indeed, using \lmone , the explicit formula
 $L_{-1} =\sum_{k=1}\infty (k+\hf)t_k\partial_{k-1}\ +z2/4$, and
differentiating, one obtains
 $${\partial u_2\over\partial\sigma}
-\sum_{k=1}\infty (k+\hf)t_k{\partial u_2\over\partial t_{k-1}}=1$$
which, on multiplying by an infinitesimal $\epsilon$, may be interpreted as the
invariance \galileo.
Clearly under a finite translation by $-\sigma$ we return to our original
equations. Thus the boundary length $L_{-1}$ and boundary cosmological constant
$\sigma$ are redundant just as was the case in ref.\bound , however
the symmetries and physical significance of our formulation are far more
transparent with $\sigma\ne0$.

Now consider 2D gravity coupled to a general $(p,q)$ minimal model. As shown
at the beginning of this section, the $W{(k)}_\alpha$
 constraints, where $\alpha$ is restricted to $-k+1\le \alpha<0$ (and $2\le
k\le q$), are missing as compared to the \hmm\ formulation. These, together
with our
 constraints \pqw , generate the full set of symmetries of the generalised KdV
hierarchy. Thus we now have spontaneous breaking
 of $\sum_{k=2}q(k-1)=\hf q(q-1)$ symmetries to which we may
 associate $\hf q(q-1)$ new parameters: $\sigma{(k)}_\alpha$.
 These parameters (and
 associated operators) will be redundant since there will be
 analogous symmetries to \galileo\ which will `gauge' them away,
 however we  again expect that they will have physical
 significance. Generalising \taup\ we have
\def\S{{\cal S}} \def\C{{\cal C}}
\eqn\taug{\tau(\sigma{(k)}_\alpha) = \S(\sigma{(k)}_\alpha ;
 W{(k)}_\alpha) \tau, \quad\quad\quad {\rm with}\ \ \S({\bf
0};W{(k)}_\alpha)=1 \ \ .}
Since the $W$-algebra of constraints is no longer a Lie algebra something more
general for $\S$ than
 $\exp\{\sum_{k,\alpha}\sigma{(k)}_\alpha  W{(k)}_\alpha\}$
may be more appropriate. The transformation \taug\ induces a similarity
transformation on the constraints $\C'=\S\C\S{-1}$ generalising \cinn , where
$\C$ runs over the $W{(k)}_n$ with $n\ge0$ and the new constraints
$\partial/\partial\sigma{(k)}_\alpha$ (which are trivially zero
 on $\tau$). The latter constraints give rise to the generalised symmetries by
similar derivations to \galileo ; Except for $L_{-1}$ these are non-local.
Since the similarity transformation preserves the
 $W$-algebra we see that the $W'{(k)}_n$ still form a
 $W$-algebra while each $\partial'_{\sigma{(k)}_\alpha}$ commutes
 with all the other constraints. It can be shown that,
by taking linear combinations of these constraints (generalising \linn ), one
can form a new simpler set of constraints of the form $(W\sigma){(k)}_n
=W{(k)}_n + corrections$ where now $n\ge1-k$ and the corrections do not
involve any $W{(k)}_m$ with $m\ge0$. It may also be shown that these
corrections can be written as a finite sum of the form $\sum_{k,\alpha}
\varsigma{(k)}_\alpha W{(k)}_\alpha$ where each
 $\varsigma{(k)}_\alpha$ is a power series expansion in all the
$\sigma{(k)}_\alpha$ with operator valued coefficients. This follows because
the original $W$-algebra with $n\ge1-k$ forms a closed algebra of constraints
under commutation.
By construction the commutator of two of these $W\sigma$
constraints also closes, however, as we will show by example below, these
constraints do not form a $W$ algebra.

In particular, consider the Ising model (or more generally coprime $p>q=3$).
In this case we have broken `symmetries' $L_{-1}$,
 $W_{-1}$ and $W_{-2}$. The most general case generalising \taup\ will involve
some combination of these three generators together
 with three parameters. The corresponding formul\ae\ to \lmone\
 will be more complicated involving mixed combinations of
 parameters and operators, because these operators do not
 commute. However, we can classify the operators under the  $Z_2$ symmetry
which flips spin (exchanges $M_+$ with $M_-$). Since
this corresponds to $d\to-d$
 (cf. comments below equation \Isingtwo ), we have $Q{1/3}\to-Q{1/3}$ and
using \lflow,  it follows that $t_r$ is $Z_2$--odd(even) if $r$ is odd(even).
 Using the explicit formul\ae\ for $W_n$ and $L_n$ (see e.g. ref.\goeree ) we
find that these are $ $ $ $ $Z_2$--odd(even)
if $n$ is odd(even). Thus we propose that $W_{-2}$ is (perhaps
 non-linearly) related to the boundary length. We do not know what
 r\^ole is played by the $Z_2$ odd $W_{-1}$. As mentioned above,
 $L_{-1}$ is the boundary magnetization. We will now confirm
 this.

Avoiding the complications of mixing let us first consider
 $L_{-1}$ in isolation.
In this case eqn.\taup\ again applies. A direct calculation of the modified
 constraints\foot{We use the conventions of ref.\goeree .}, via the method
  of \direct , gives: 
\eqn\wsig{ W\sigma_n =W_n-(n+2)\sigma{n+1}W_{-1}+(n+1)\sigma{n+2}W_{-2}
\hskip 1.5cm n\ge0}
with the $L\sigma_n$, $n\ge-1$, as before. It is clear (by counting) from the
 discussion below \taug\ that there are no $W\sigma_{-1}$ or
 $W\sigma_{-2}$ constraints; Indeed it is amusing to note that this is already
incorporated in \wsig\  if we take $n\ge-2$.
This is preserved by the modified algebra of constraints which by explicit
computation is found to be:

\def\ls#1 {L\sigma_{#1}}
\def\ws#1 {W\sigma_{#1}}
\def\us#1 {U\sigma_{#1}}
$$\eqalign{
[\ls n , \ls m ] &=(n-m)\ls n+m \cr
[\ls n ,\ws m ] &=(2n-m)\ws n+m -(2n+1)(m+2)\sigma{m+1}\ws n-1 \cr
                      & +2(n+1)(m+1)\sigma{m+2}\ws n-2 \cr
[\ws n , \ws m ] &=
-{1\over3}(n-m)\{(n2+4nm+m2)+9(n+m)+14\}\ls n+m \cr
               & +2(n-m)(\us n+m +\sigma{n+m+3}\us -3 )\cr
&+\left\{ {1\over3}(n+1)(m+2)(m+1)m\sigma{n+2}\ls m-2 \right.\cr
&-{1\over3}(n+2)(m+3)(m+2)(m+1)\sigma{n+1}\ls m-1 \cr
&+2(n+2)(m+1)\sigma{n+1}\us m-1 -2(m+2)(n+1)\sigma{n+2}\us m-2 \cr
&\ \ \left.-(n\leftrightarrow m)\right\}\cr
}$$
where $\us n $ may be taken to be $\sum_{k\le-2}\ls k \ls n-k
+\sum_{k\ge-1} \ls n-k \ls k $. The corresponding (galilean) symmetry follows,
cf. \galileo ,
 from equation \Galileo:
$$
\eqalign{
 u_3 &\rightarrow u_3+\epsilon \cr
 u_2 &\rightarrow u_2\cr
\s &\rightarrow \s+\epsilon\cr
z &\rightarrow z+\epsilon{4\over 3}t_{1,1} \cr
\B &\rightarrow \B+\epsilon{5\over 3}t_{2,1}\cr
 \tlk &\rightarrow \tlk -\epsilon(k+1+{l\over 3})t_{l,k+1} \hskip 2.5cm k \ge
0 ;l=1,2
}
$$
and the string equation, from the $L'_0$ constraint, is the one in \strngtwo.
Since it is $u_3$ that shifts by a constant we
 identify $\sigma$, by arguments similar to those below \vircon
 , as the boundary magnetic field and
$L_{-1}$ as the boundary magnetization\bound .

Finally consider $W_{-2}$. If we take $\tau(\theta) =\exp(\theta W_{-2})\tau$
then the corrections to the constraints \pqw\ can be computed perturbatively
in $\theta$. We find: 
\def\lt#1 {L\theta_{#1}}
\def\wt#1 {W\theta_{#1}}
\def\l#1 {L_{#1}}
\def\w#1 {W_{#1}}
$$\eqalign{\lt0 &= \l0 -2\theta{\partial\over\partial\theta}\cr
\lt1 &= \l1 -4\theta \w-1 +8\theta2\l-2 \l-1 +\cdots\cr
\lt2 &= \l2 -12\theta2\l-1 2+\cdots\cr
\lt3 &= \l3 -16\theta2\l-1 +\cdots\cr
\wt-2 &= \w-2 -{\partial\over\partial\theta}\cr
\wt0 &= \w0 -4\theta\l-1 2-8\theta2(\l-2 \w-2 +2\l-3 \w-1 )+\cdots\cr
\wt1 &= \w1 -4\theta\l-1 -6\theta2 (\l-1 \w-2 +2\l-2 \w-1 )+\cdots\cr
\wt2 &= \w2 -8\theta2 (3\w-2 +2\l-1 \w-1 )+\cdots\cr}$$
The remaining (positive $n$) constraints receiving no corrections to order
$\theta2$. The commutation algebra is evidently even less illuminating.
The corresponding symmetry follows from the explicit formula for $W_{-2}$ and
is highly non-local.

\bigskip
\bigskip
\bigskip
\bigskip
\noindent
{\bf Acknowledgements.}

B.S. would like to thank Ken Barnes for arranging
 his tenure at Southampton during the Summer
of 1991. T.R.M and C.J. thank Simon Dalley for helpful comments.
C.J. and B.S. are grateful to  the S.E.R.C. for financial support.

\listrefs
\vfill\eject
\noindent
\centerline{\bf Figure Captions.}
\bigskip
\bigskip
\bigskip
\noindent
{\bf Figure 1(a):} The numerical solution to the
Ising model $\pqq$ string equation \Isingfour\
for the string susceptibility $\r(z)$, with asymptotics
$\r(\infty)=z{1/3}$ and $\r(-\infty)=0$.
The integration was
performed on a mesh of 1600 points
over a range $z=\pm 200$, with a maximum error of $10{-5}$.
Here, $z$ is the cosmological constant.

\bigskip
\noindent
{\bf Figure 1(b):} The numerical solution to the Lee--Yang model
$\pqq$ string equation (equation \smiley\ with $t_k=16/35\deltak_3$)
for $\r(z)$, with asymptotics
$\r(\infty)=z{1/3}$ and $\r(-\infty)=0$. An identical integration procedure
to that used for the Ising model string equation was employed.

\bye